\documentclass{article}

\usepackage{arxiv}

\usepackage[utf8]{inputenc} 
\usepackage[T1]{fontenc}    
\usepackage{hyperref}       
\usepackage{url}            
\usepackage{booktabs}       
\usepackage{amsmath}        
\usepackage{amsfonts}       
\usepackage{nicefrac}       
\usepackage{microtype}      
\usepackage{lipsum}		
\usepackage{graphicx}
\usepackage{subcaption}     
\usepackage[authoryear,round]{natbib}
\usepackage{doi}

\newcommand{\rev}[1]{#1}

\newcommand{\shortcite}[1]{\citep{#1}}
\newcommand{\shortciteN}[1]{\citet{#1}}
\newcommand{\shortciteANP}[1]{\citeauthor{#1}}

\title{Simulating Cognitive Smart Freight Corridors with Agent-Based Models and Reinforcement Learning}

\author{ \href{https://orcid.org/0000-0001-7334-9500}{\includegraphics[scale=0.06]{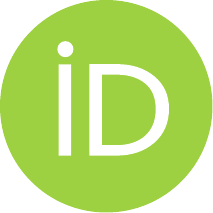}\hspace{1mm}Madelaine Martinez-Ferguson}\thanks{Corresponding author}\\
	Department of Industrial and Systems Eng.\\
	University of Tennessee, Knoxville\\
	Knoxville, TN, USA, 37996 \\
	\texttt{mmart199@vols.utk.edu} \\
	\And
	\href{https://orcid.org/0000-0001-6994-0146}{\includegraphics[scale=0.06]{orcid.pdf}\hspace{1mm}Chun Wang} \\
	Concordia Institute for Information Systems Eng.\\
	Concordia University\\
	Montreal, QC, CANADA \\
	\texttt{chun.wang@concordia.ca} \\
	\AND
	\href{https://orcid.org/0000-0001-7465-7783}{\includegraphics[scale=0.06]{orcid.pdf}\hspace{1mm}Mustafa Can Camur} \\
	Amazon.com, Inc. \\
	New York,  NY, USA, 10001 \\
	\texttt{mccamur@amazon.com} \\
	\And
	\href{https://orcid.org/0000-0003-1990-0159}{\includegraphics[scale=0.06]{orcid.pdf}\hspace{1mm}Xueping Li} \\
	Department of Industrial and Systems Eng.\\
	University of Tennessee, Knoxville\\
	Knoxville, TN, USA, 37996 \\
	\texttt{Xueping.Li@utk.edu} \\
}

\renewcommand{\shorttitle}{Simulating Cognitive Smart Freight Corridors}

\hypersetup{
pdftitle={A template for the arxiv style},
pdfsubject={q-bio.NC, q-bio.QM},
pdfauthor={David S.~Hippocampus, Elias D.~Striatum},
pdfkeywords={First keyword, Second keyword, More},
}

\begin{document}
\maketitle

\begin{abstract}
Smart freight corridors offer a practical pathway for connected and automated vehicle (CAV) deployment in freight transportation, but physical experimentation is expensive and existing approaches rely on predefined control policies that cannot capture adaptive behaviors. This paper presents an agent-based modeling (ABM) framework coupling a physical infrastructure layer, a connectivity layer (V2X), and a decision layer integrating reinforcement learning (RL) and multi-agent reinforcement learning (MARL) for platoon formation and charging coordination. We evaluate three scenarios (Baseline, Assisted, and Cognitive) using throughput, congestion, energy, emissions, and robustness metrics. Preliminary results indicate that the Cognitive scenario achieves higher throughput and lower congestion than the baseline, while the Assisted scenario delivers meaningful energy savings per kilometer through platooning. Sensitivity analysis shows that the throughput advantage of the smart corridor widens under conditions with high demand and that MARL coordination extracts greater utilization from fixed charging capacity than rule-based assignment.
\end{abstract}


\section{INTRODUCTION} \label{sec:introduction}


The U.S. trucking industry is the backbone of domestic freight movement, carrying 11.27 billion tons in 2024 and accounting for 72.6\% of total domestic freight tonnage \shortcite{ata2024trends}. 
Traffic congestion poses a severe and growing threat to the cost efficiency of this sector; in 2022 alone, corridor congestion cost the trucking industry \$108.8 billion in lost productivity \shortcite{short2024congestion}. CAVs, vehicles that combine onboard sensing and machine learning with vehicle-to-vehicle (V2V), vehicle-to-infrastructure (V2I), and broader vehicle-to-everything (V2X) communication, offer a promising means to address this challenge, demonstrating potential to dampen traffic flow instabilities and reduce fuel consumption through cooperative platooning, with energy savings of up to 25\% reported under controlled conditions \shortcite{stern2018dissipation}.  
However, the magnitude of these benefits depends nonlinearly on penetration rates, infrastructure design, and the control policies governing corridor interactions \shortcite{gueriau2020quantifying}, making technology at the vehicle level insufficient on its own.

Realizing the full potential of CAVs requires supporting infrastructure. Smart freight corridors, dedicated highway segments equipped with intelligent transport systems (ITS) such as sensors and V2X communication, provide this environment by concentrating investments on high-impact links while allowing controlled operational policies to be enforced. Active deployments such as the I-24 Smart Corridor in Tennessee, which has deployed a V2X roadmap along a major freight artery \shortcite{hill2025i24}, illustrate the practical urgency of this direction. Yet deploying such corridors effectively raises fundamental system-level questions \shortcite{soto2022survey}: what connectivity and automation capabilities are needed, which control policies deliver measurable improvements, and how robust corridor operations remain under incidents and demand surges.

These decisions span three interdependent layers. The physical layer comprises the corridor infrastructure, the cyber layer handles V2X communication, and the decision layer governs operational aspects such as platoon formation policies and dynamic pricing. \rev{Coordinating them jointly is analytically intractable under stochastic freight demand, heterogeneous vehicle capabilities, and dynamic V2I interactions \shortcite{aslam2025intelligent}, and} real-world pilots cannot cover the full combinatorial space of infrastructure configurations, disruption scenarios, and control policies \shortcite{lam2014electric}. \rev{Moreover, existing simulation models address individual aspects, such as RL for platoon coordination or charging control, but none captures the joint adaptive behavior across these layers, leaving a critical gap in the assessment of smart corridor operations.} 

This paper addresses that gap through four contributions. First, we develop a modular Python-based \rev{ABM} environment for smart freight corridors with a three-layer architecture covering physical, cyber, and decision components. Second, we integrate single-agent RL for corridor-level platoon formation and lane management, and MARL for distributed charging station coordination. \rev{The corridor controller selects among five discrete actions (no operation, encourage platoons, dissolve platoons, activate managed lanes, deactivate managed lanes); each of the five station agents independently selects among five pricing and priority actions (maintain price, increase price, decrease price, enable priority, disable priority) informed by neighbor state via two communication rounds per step.} Third, we evaluate three scenarios (Baseline, Assisted, Cognitive) across key performance metrics including throughput, congestion, energy consumption, emissions, and disruption robustness. Fourth, we quantify the value of cognitive corridor control relative to the unassisted baseline. The remainder of this paper is organized as follows. Section~\ref{sec:related-work} reviews related work. Section~\ref{sec:methodology} presents the corridor conceptual model, simulation implementation, and RL/MARL formulation. Section~\ref{sec:results} details experimental design and results. Section~\ref{sec:conclusions} concludes with implications and future research.

\section{RELATED WORK} \label{sec:related-work}
We position this work at the intersection of \rev{two research streams: smart freight corridors, and learning methods for corridor control. Key contributions and gaps are discussed below.}

\subsection{Smart Freight Corridors} \label{subsec:smart-freight}

\rev{Smart freight corridors integrate a physical layer of road segments, managed lanes, and charging stations with a cyber layer of V2X communication, sensing, and real-time data exchange, forming cyber-physical systems (CPS) that extend traditional highway infrastructure. In characterizing these systems, \shortciteN{pompigna2022smart} identify four core functions, including V2X connectivity, self-adaptability, and energy harvesting, that collectively enable ITS applications such as variable speed limits and cooperative adaptive cruise control. \shortciteN{wu2025digital} operationalize this coupling through a digital twin framework for a cellular-V2X-enabled corridor that replicates vehicle behaviors, signal timing, and traffic patterns in a virtual environment synchronized with the physical system.}

\rev{Building on this infrastructure, several studies optimize CAV coordination across the physical and cyber layers at corridor scale. \shortciteN{yu2019corridor} formulate a mixed-integer linear programming model to optimize CAV trajectories along a signal-free corridor, capturing car-following and lane-changing interactions to minimize total vehicle delay. Expanding the scope from trajectory control to vehicle grouping, \shortciteN{sakaguchi2023cyber} develop a cyber-physical framework that organizes CAVs into platoons and applies receding horizon optimization for lane coordination on multi-lane freeways, achieving real-time performance with gains in fuel economy and travel time. \shortciteN{peng2025enhancing} further incorporate mixed-traffic conditions, combining a platoon catch-up mechanism with dedicated lane management under varying CAV penetration rates and showing that the hybrid control increases road capacity and reduces shockwave propagation. At the modeling level, \shortciteN{palmieri2023co} extend CPS representation through co-simulation, decomposing platoon dynamics to simplify testing under varying conditions (e.g., adverse weather).}

\rev{Beyond framework development, system-level assessments confirm that corridor management yields substantial returns. Using a multi-criteria method, \shortciteN{mbiydzenyuy2018impact} finds that managing traffic flow and speed is the most impactful lever for capacity utilization, with corridor section interventions generating the highest socioeconomic returns among all evaluated ITS use cases. For electric freight specifically, charging availability along a corridor measurably affects truck throughput and delays, making charging coordination a non-trivial operational decision \shortcite{gonzalez2025impact}. These studies demonstrate that the physical, cyber, and decision layers of smart freight corridors have been integrated to varying degrees, with coordination across them improving corridor performance. However, the decision layer remains limited to predetermined optimization models or fixed control rules that do not adapt to stochastic demand, disruptions, or evolving conditions. This motivates the integration of adaptive learning methods, reviewed next.}

\subsection{Simulation and Learning Methods for Corridor Control} \label{subsec:simulation-learning}

\rev{Some prior frameworks do integrate elements of the physical, cyber, and decision layers, though with important limitations. \shortciteN{liu2026robust} extend this line of work by introducing a DRL agent that uses real time signal phase and timing data to support driving decisions for a CAV at the corridor level. Their results show that adaptive learning can operate across layers. However, their framework focuses on optimizing the energy use of a single vehicle trajectory, rather than coordinating multiple corridor operations.}

\rev{Scaling adaptive control to multi-agent settings, RL and MARL have been applied to individual corridor decisions with promising results. \shortciteN{han2025improved} apply a centralized-training, decentralized-execution (CTDE) MARL scheme to jointly coordinate ramp metering and variable speed limits along a freeway corridor, optimizing mobility, safety, and emissions simultaneously. Their work demonstrates that MARL can learn coordinated control policies across multiple corridor sections, but targets passenger vehicle traffic without freight agents, charging coordination, or disruption scenarios. \shortciteN{li2025real} apply the same CTDE paradigm to EV charging station control, where each charger acts as an independent agent learning dynamic power allocation policies under stochastic arrivals and departures. While this confirms that MARL can effectively manage charging decisions, the framework is confined to a single station without corridor dynamics, platoon behavior, or freight context.}

\rev{In the freight domain, \shortciteN{barba2025real} train a proximal policy optimization agent to jointly plan routes and charging stops for electric trucks under stochastic queuing and electricity pricing, showing that RL can address charging complexity specific to freight operations. However, the approach operates at the network routing level with a single agent, without modeling corridor infrastructure as an active control layer. \shortciteN{alam2026network} address the problem most closely related to ours by jointly optimizing charging schedules and platooning decisions for electric trucks at the network scale. Their work confirms the operational value of coordinating these decisions for electric freight, but formulates the problem as a routing optimization model rather than an adaptive learning framework, without representing heterogeneous agent interactions through ABM or incorporating disruption scenarios. Taken together, these studies demonstrate the value of individual components but leave a critical gap. To the best of our knowledge, no existing framework simultaneously integrates ABM to represent heterogeneous freight agents, treats corridor infrastructure enabled by V2X as an active control layer, and learns joint policies for platoon formation, charging coordination, and disruption response. This work addresses that gap.}

\section{METHODOLOGY} \label{sec:methodology}
This section details our simulation framework: the smart freight corridor conceptual model, implementation, and RL/MARL formulations for adaptive control.

\subsection{Corridor Conceptual Model} \label{subsec:conceptual-model}
\begin{figure}[htbp]
    \centering
    \rev{\includegraphics[width=0.85\columnwidth]{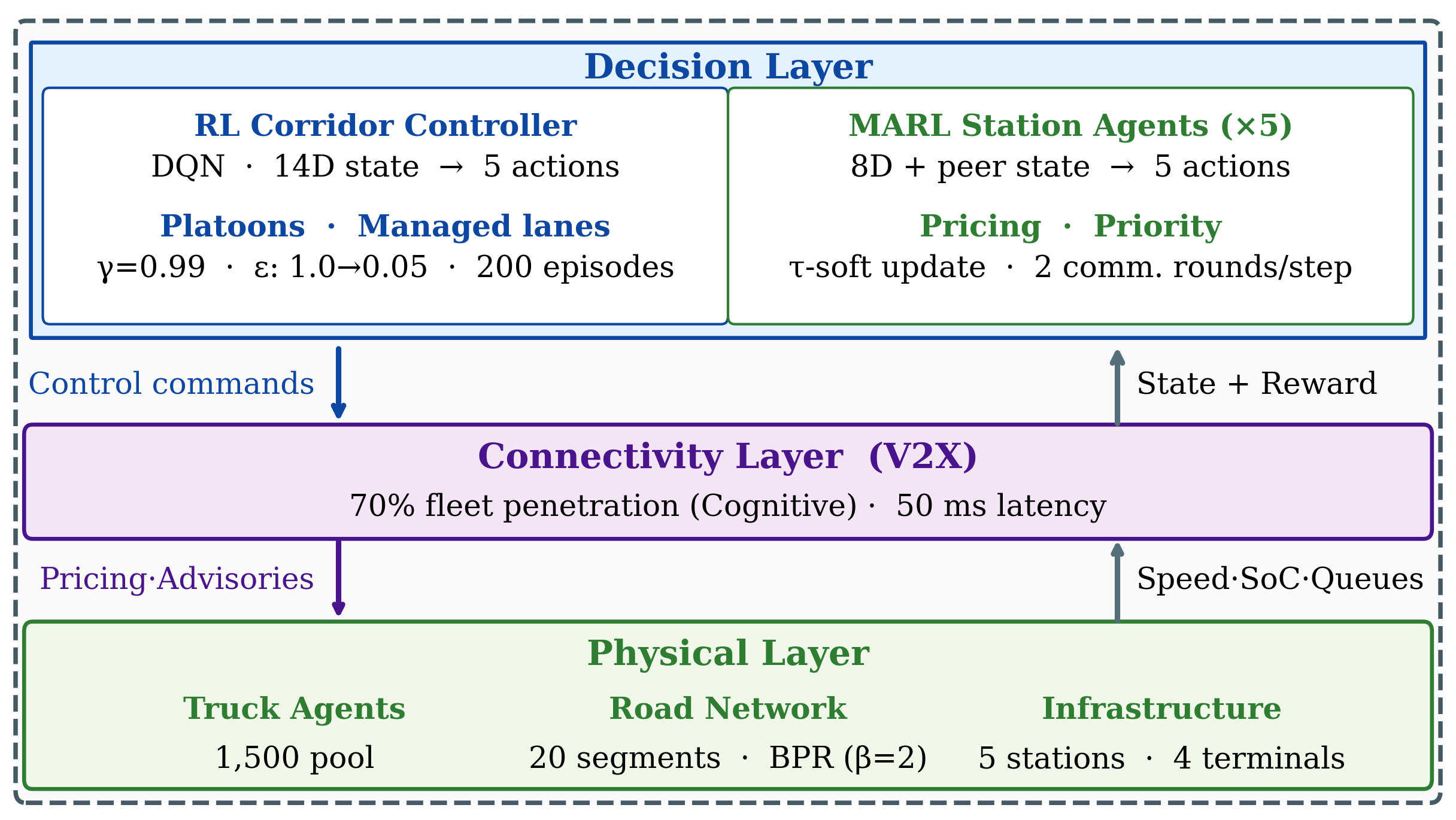}
    \caption{Three-layer architecture of the smart freight corridor, showing agent types, information exchanged, and policy outputs at each layer.}
    \label{fig:architecture}}
\end{figure}

We conceptualize the smart freight corridor as a three-layer system integrating physical infrastructure, cyber connectivity, and adaptive decision making (Figure~\ref{fig:architecture}). 

The physical layer comprises the tangible infrastructure elements. Road segments are characterized by length, capacity, speed limits, and number of lanes, with varying conditions due to disruptions. \rev{Each charging station along the corridor has three ports with queue management and power levels of 150, 250, or 350 kW, reflecting current heavy-duty charging practice of serving up to four vehicles with 35--60 minute sessions \shortcite{harris2023review}.} Freight terminals serve as origin-destination nodes that generate demand and process arrivals. When activated by the RL controller, a managed lane opens corridor-wide across all segments, providing autonomous trucks with an uninterrupted priority lane for platooning without encountering regular-lane congestion at any point along the corridor. Traffic flow on segments follows the \shortciteANP{bpr1964} function (BPR):
\begin{equation}
    t_e = t_e^0 \left(1 + \alpha \left(\frac{f_e}{C_e}\right)^{\!\beta}\right),
    \label{eq:bpr}
\end{equation}
where $t_e^0$ is free-flow travel time, $f_e$ is flow, $C_e$ is capacity, and $\alpha=0.15$, $\beta=2$ are calibration parameters tuned for freight-dominated corridors. \rev{We depart from the standard $\beta=4$, known to overestimate travel times under congestion \shortcite{kucharski2017estimating}. Freight-dominated streams exhibit more gradual congestion onset due to longer headways and limited heavy vehicle acceleration, making $\beta=2$ physically realistic for this corridor type.} The congestion ratio $f_e/C_e$ is left uncapped so that congested segments yield the full BPR penalty, and segment flow is computed using free-flow speed.

\rev{The connectivity layer enables information sharing via V2X communication encompassing V2I and V2V messaging with configurable penetration rates and latency. The penetration rate denotes the fraction of trucks equipped with active V2X transceivers (100\% among V2X capable trucks in the Assisted and Cognitive scenarios; thus, effective fleet level penetration equals the scenario CAV share of 70\%); latency is modelled as a fixed 50~ms per-message delay, within the standard V2X communication bounds \shortcite{Kenney2011DSRC}. Connected agents receive real-time data on traffic speeds, incidents, queue lengths, and charging availability, alongside broadcast operational constraints covering policy rules, safety envelopes, and service priorities. Unlike prior models assuming all-or-nothing connectivity, the explicit cyber layer allows penetration rate to be varied as an experimental parameter, separating infrastructure capability from control policy. This enables three scenario tiers: Baseline assumes no connectivity, Assisted adds V2X and supervised automation, and Cognitive further incorporates RL and MARL optimization.}

\rev{The decision layer implements decentralized control via two agent types: an RL corridor controller governing platoon formation and lane activation, and five MARL station agents managing local pricing and priority actions (see Section~\ref{subsec:rl-marl} for details).}

\subsection{Simulation Implementation} \label{Subsection:SimulationImplementation}
\begin{figure*}[htbp]
    \centering
    \includegraphics[width=0.85\textwidth]{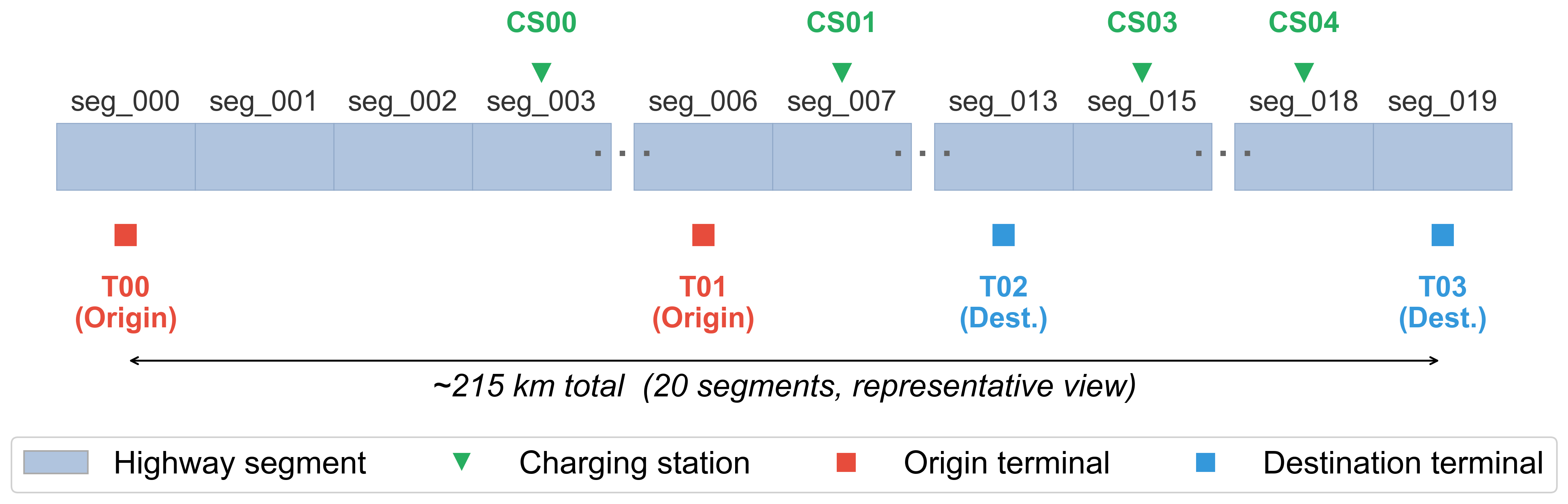}
    \caption{Layout of the simulated corridor network (seed~42 instance).}
    \label{fig:corridor_network}
\end{figure*}

We implement the corridor simulation as a modular Python-based ABM \shortcite{bonabeau2002abm} with the five agent types organized in their respective tiers (Figure~\ref{fig:corridor_network}). \rev{ABM is used instead of other simulation approaches as heterogeneous autonomy modes and individual SoC trajectories can be represented directly through agent rules. This structure allows all three autonomy modes to coexist within each replication, capturing mixed autonomy corridor dynamics without separate runs by vehicle type.} 

Each freight truck agent $i$ maintains state $s_i(t) = \langle v_i, \text{SoC}_i, \kappa_i, p_i \rangle$ where $v_i \in \mathcal{V}$ denotes current segment location, $\text{SoC}_i \in [0,1]$ represents battery state-of-charge, $\kappa_i \in \{\text{manual}, \text{assisted}, \text{autonomous}\}$ indicates autonomy mode, and $p_i$ captures platoon membership and role (leader, follower, or none). Truck behaviors include route following, platoon join/leave decisions, charging requests, and compliance with corridor advisories. Energy consumption follows $E_i = r_{\text{base}} \cdot d_i$ for solo driving and $E_i = r_{\text{base}} \cdot (1 - \eta_p) \cdot d_i$ for platoon followers, where $r_{\text{base}} = 1.2$ kWh/km is base consumption and $\eta_p = 0.15$ represents platooning energy savings. \rev{Both parameters are empirically grounded: $r_{\text{base}}$ reflects measured heavy-duty electric truck consumption \shortcite{barba2025real}, and $\eta_p$ is consistent with reported fuel reductions of 10--25\% under cooperative platooning conditions \shortcite{tsugawa2016platoon}.}

The corridor control agent aggregates corridor state into a 14-dimensional observation vector:
\begin{equation} \label{eq:obs}
    \mathbf{o}_{\text{corridor}} = [\bar{c}, c_{\max}, n_{\text{inc}}, \bar{u}_s, q_{\text{total}}, r_p, \overline{\text{SoC}}, n_{\text{charge}}, \mathbf{f}_{\text{scen}}, \tau, m_{\text{open}}, \bar{c}_{m}],
\end{equation}

where $\bar{c}$ is average congestion, $c_{\max}$ is maximum congestion, $n_{\text{inc}}$ is incident count, $\bar{u}_s$ is average station utilization, $q_{\text{total}}$ is total queue length, $r_p$ is platoon rate, $\overline{\text{SoC}}$ is fleet average SOC, $n_{\text{charge}}$ is trucks needing charging, $\mathbf{f}_{\text{scen}} \in \{0,1\}^3$ is a scenario flag vector (Cognitive, Assisted, Connected), $\tau$ is normalized simulation time, $m_{\text{open}} \in \{0,1\}$ indicates whether managed lanes are currently open, and $\bar{c}_{m}$ is the average congestion on managed-lane segments. 

The simulation advances in discrete time steps of $\Delta t = 0.08$ hours (5 minutes). Each step proceeds through five phases: controllers select actions based on current observations; traffic flow updates based on truck positions and segment capacities; trucks update positions, consume energy, and make decisions; charging processes and V2X state update; and metrics are recorded for analysis.

\subsection{RL/MARL Formulation} \label{subsec:rl-marl}
We formulate corridor control as RL problems, enabling adaptive policies that improve through experience. The corridor control problem is posed as a Markov Decision Process \shortcite{bellman1957mdp} for the central controller and a Markov game for distributed station agents. The corridor state representation consists of:
\begin{equation} \label{eq:state}
S(t) = \langle \{s_i(t)\}_{i \in \mathcal{T}}, \mathbf{q}(t), \mathbf{C}(t), \mathbf{d}(t) \rangle,
\end{equation}
where $\mathbf{q}(t)$ captures charging station queues, $\mathbf{C}(t)$ represents link capacities, and $\mathbf{d}(t)$ indicates disruption status. The optimization objective balances multiple key performance indicators:
\begin{equation} \label{eq:objective}
J = \mathbb{E}\left[\sum_{t=0}^{T} \gamma^t \big(\alpha \cdot \text{Throughput}(t) - \beta \cdot \text{Delay}(t) - \delta \cdot \text{Energy}(t) - \eta \cdot \text{Instability}(t)\big)\right],
\end{equation}
\rev{where $\gamma = 0.99$ reflects the long-horizon nature of corridor control, where platoon and charging decisions at one step propagate through queue and SoC dynamics for many steps. Each episode spans 125 steps at $\Delta t = 0.08$~h (5-minute resolution, 10 simulated hours), covering a representative freight operating shift. The replay buffer of 25{,}000 transitions covers the full training horizon ($200 \times 125 = 25{,}000$ steps), ensuring uniform sampling across exploration and exploitation phases; the buffer reaches capacity by episode 80, after which gradient updates increasingly reflect recent experience. Learning rate decay (rate 0.98, floor $10^{-4}$) satisfies the Robbins--Monro convergence condition for stochastic gradient descent, reaching the minimum threshold near episode 114.}


\rev{This architecture reflects a principled decomposition: corridor-level decisions require global state and benefit from a single coordinating policy, while charging coordination is inherently distributed, making independent MARL with communication more appropriate. The multi-objective reward, equation~(\ref{eq:objective}), jointly optimizes throughput, delay, energy, and stability, unlike single-objective corridor RL formulations in prior work.} The corridor controller implements this global policy via a DQN-style agent \shortcite{mnih2015dqn}, operating over a 14-dimensional state space,  capturing congestion statistics, station utilization, queue length, platoon rate, fleet SoC, charge demand, managed lane status, and scenario flags as in equation~(\ref{eq:obs}). The agent selects among five discrete actions governing platoon formation and lane activation, with a reward combining energy savings, throughput improvement, congestion penalty, and incident penalty. The Q-network uses a two-layer architecture with 64 hidden units and ReLU activation, trained with experience replay (buffer size 25,000) and target network updates every 100 steps.

In contrast, charging coordination is handled through MARL, where each of the five station agents operates independently with an 8-dimensional local state capturing utilization, queue length, port availability, operational status, pricing, and service metrics. Each agent selects from five pricing and priority actions, \rev{with price adjustments applied as multiplicative steps of $\pm$10\% per action, enabling gradual price evolution over successive steps without requiring a large action space.} Agents broadcast their full local state across two communication rounds per step, with a reward balancing local objectives (minimizing wait time, optimizing utilization) against global load distribution. Soft target updates with $\tau = 0.01$.

\subsection{Scenario Definitions} \label{Subsection:ScenarioDefinitions}
We define three scenarios that represent progressive levels of connectivity, automation, and adaptive intelligence applied to the same physical corridor infrastructure. \rev{The 70\% CAV penetration adopted here lies in the range where platoon control and lane management yield significant capacity and stability improvements \shortcite{peng2025enhancing}.}

\textbf{Baseline.} All trucks operate in fully manual mode ($\kappa_i = \text{manual}$, 100\%). No V2X communication is active; trucks follow fixed routes, and select charging stops based on individual SOC thresholds. This scenario serves as the performance reference for all comparisons.

\textbf{Assisted.} The cyber layer is fully enabled: V2I and V2V messaging provide real-time traffic conditions, incident alerts, and charging availability. Seventy percent of trucks operate in assisted mode ($\kappa_i = \text{assisted}$, SAE L2/L3) and 30\% remain manual. Platooning follows a rule-based formation policy; charging coordination uses a greedy least-loaded assignment. No learning agents are active.

\textbf{Cognitive.} The full three-layer architecture is active. Seventy percent of trucks operate autonomously ($\kappa_i = \text{autonomous}$, SAE L4/L5) and 30\% in assisted mode. The corridor controller employs the trained DQN policy for platoon advisories and the five station agents apply trained MARL policies for dynamic pricing and priority assignment; during evaluation agents execute greedy policies with $\epsilon = 0.05$.

\section{PRELIMINARY RESULTS AND DISCUSSION} \label{sec:results}

\subsection{Experimental Design} \label{subsec:experimental-design}
We evaluate the three scenarios defined in Section~\ref{Subsection:ScenarioDefinitions} (Baseline, Assisted, and Cognitive) using the corridor network described in Section~\ref{Subsection:SimulationImplementation}. All three scenarios share identical physical infrastructure for a given random seed, ensuring fair comparison.

We measure corridor performance across six dimensions: throughput (completed trips per hour), travel time (average trip duration in hours), energy consumption (kWh per kilometer), emissions proxy (CO$_2$ emissions based on energy and grid carbon intensity of 0.4 kg/kWh), congestion index (average ratio of flow to capacity across segments), and robustness score (ratio of disrupted to baseline performance).

Experiments use 1,500 trucks, 20 road segments, 5 charging stations, and 4 terminals. Each segment length is drawn uniformly from $[10, 15]$~km, yielding corridor lengths that vary across replications ($200$--$300$~km for 20 segments, with a mean of $\sim$250~km). Importantly, for a given random seed, the corridor layout is identical across all scenarios, ensuring a fair comparison. Episode length is 125 steps (10 simulated hours, at $\Delta t = 0.08$ hours per step). Scenario comparison runs 5 replications per scenario; disruption tests run 5 replications per condition. RL training runs \rev{200} episodes with 125 steps each. \rev{Given the stochastic corridor generation, seed-to-seed variation in layout contributes substantially to metric variance; with five replications, results should be interpreted as directional and preliminary, consistent with the scope of this paper.}
\rev{To assess real-time applicability, inference latency was benchmarked over 10,000 forward passes: the DQN corridor controller requires 3.8~$\mu$s per action and the five MARL station agents complete two communication rounds in 42.9~$\mu$s, giving a combined decision latency of 47~$\mu$s per simulation step, four orders of magnitude below the 5-minute (300,000~ms) control interval. Once trained, the policy can issue corridor and pricing decisions in real time with negligible computational overhead, making it compatible with operational deployment on standard edge hardware.}

\subsection{RL Training Convergence} \label{subsec:rl-training}

\begin{figure}[h!]
    \centering
    \begin{subfigure}[h]{0.48\columnwidth}
        \raggedright
        \includegraphics[width=0.9\textwidth]{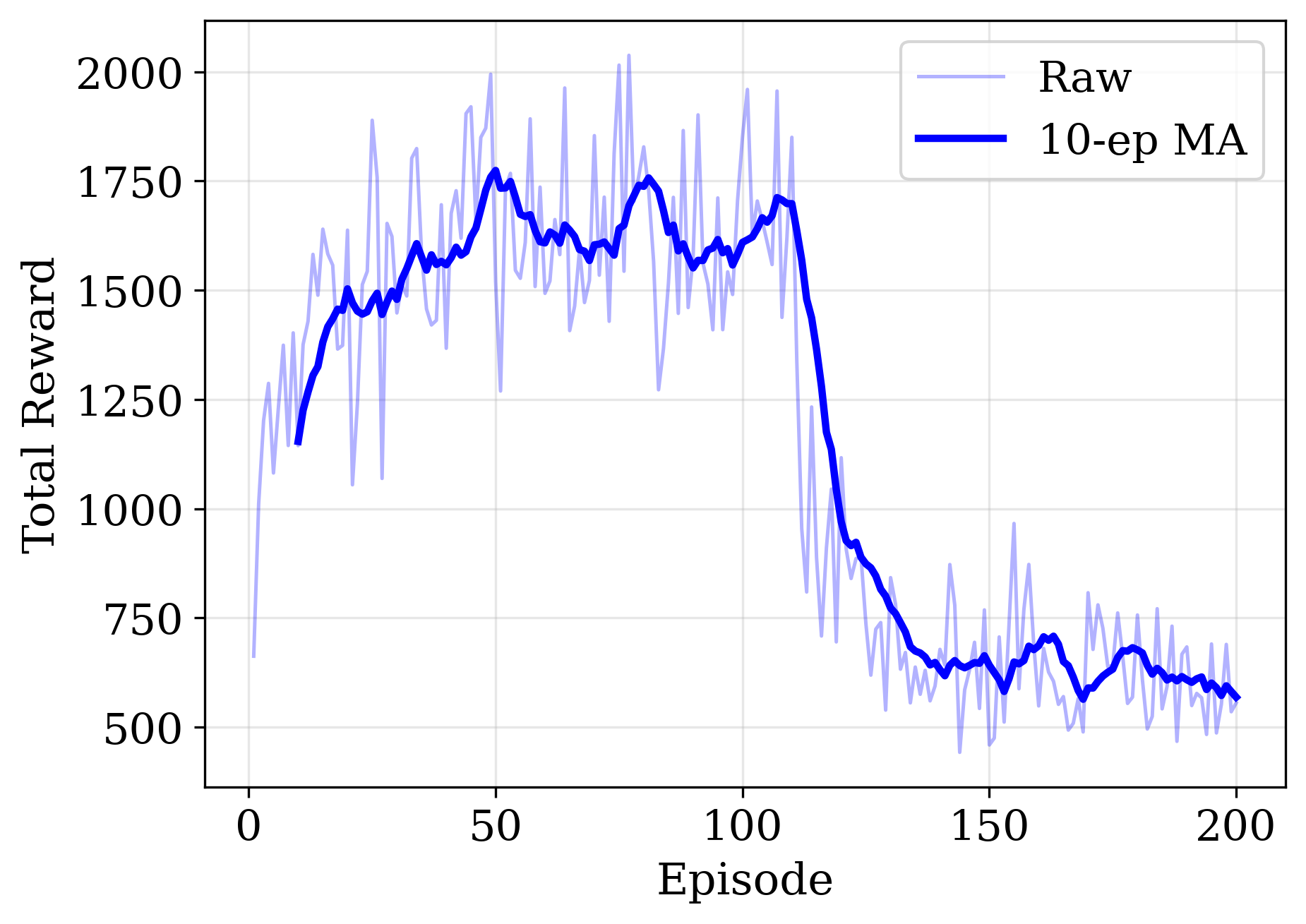}
        \raggedright
        \caption{Total reward (corridor + station).}
        \label{fig:learning_reward}
    \end{subfigure}
    \hfill
    \begin{subfigure}[h]{0.48\columnwidth}
        \centering
        \includegraphics[width=0.9\textwidth]{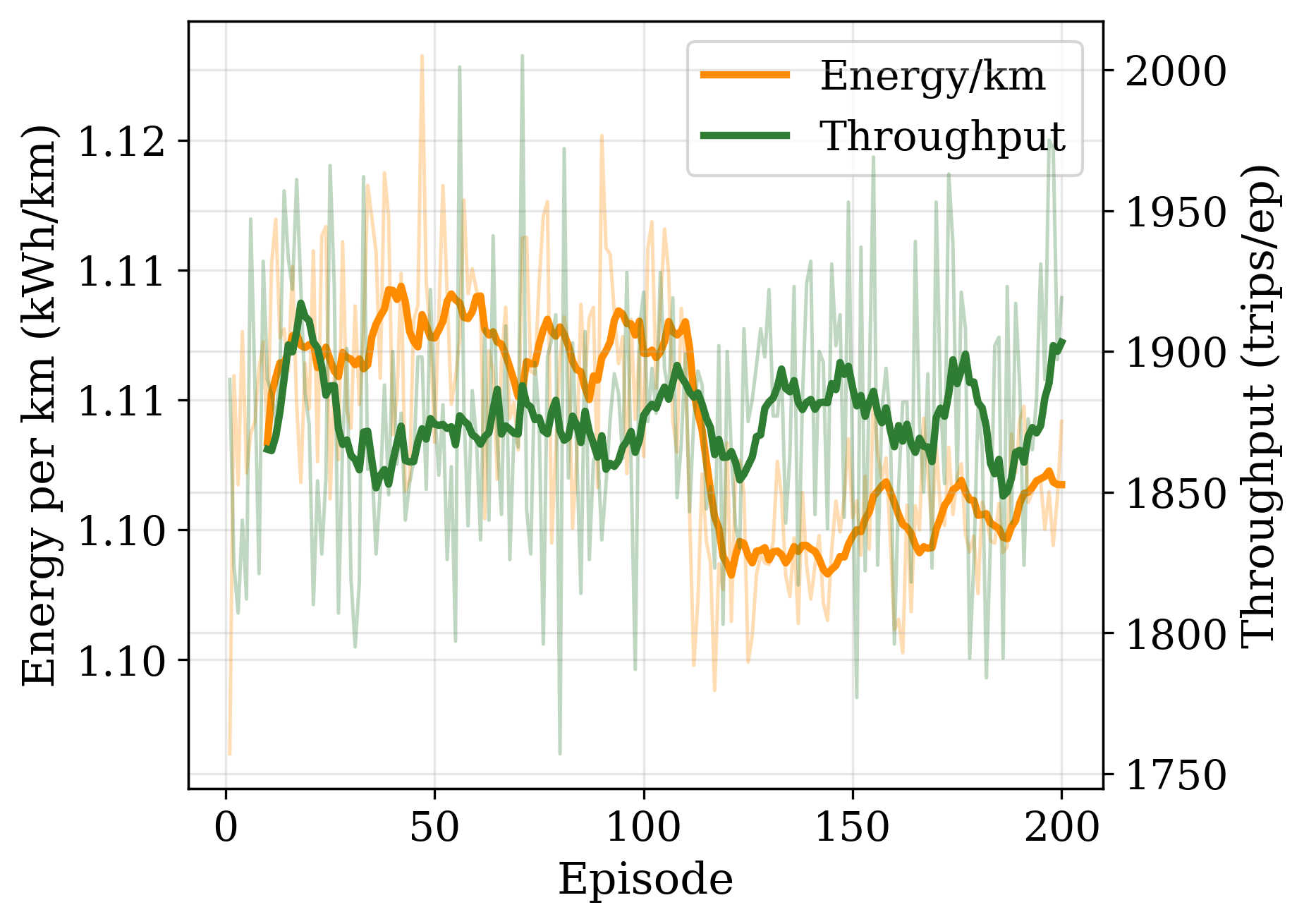}
        \caption{Energy efficiency and throughput.}
        \label{fig:learning_energy}
    \end{subfigure}
    \caption{RL/MARL training curves for the Cognitive scenario.}
    \label{fig:learning}
\end{figure}

Figure~\ref{fig:learning} shows RL/MARL training dynamics for the Cognitive scenario across \rev{200 episodes. 
Figure~\ref{fig:learning}\subref{fig:learning_reward} shows the combined reward (corridor + station) ranging from 442 to 2,038 (best: 2,038 at episode~77), with rewards rising through the first 80 episodes before declining as $\varepsilon$ approaches its minimum; the MARL station component improves from $-$1,378 at episode~1 to $-$386 at episode~200 (+72\%), reflecting progressive coordination learning. Figure~\ref{fig:learning}\subref{fig:learning_energy} shows energy per kilometer holding near 1.11~kWh/km throughout, confirming the agent does not trade energy efficiency for reward, alongside corridor throughput which stabilizes at $\sim$1,876 completed trips per episode (range: 1,757--2,005).} 


\subsection{Scenario Comparison and Sensitivity Analysis} \label{subsec:scenario-comparison}

For scenario comparison, each scenario runs a 10-hour horizon (125 steps × 0.08 h) over a 1,500-truck fleet pool with five replications, with the autonomy mix described in Section~\ref{Subsection:ScenarioDefinitions}. Table~\ref{tab:scenario_metrics} presents performance across the three scenarios averaged over five replications. The Cognitive scenario delivers the largest gains, achieving \rev{27.0\%} higher throughput and a \rev{63.9\%} reduction in the congestion index relative to Baseline, driven by the RL corridor controller actively managing lane access and platoon policies. Energy consumption per kilometer improves under both Assisted and Cognitive scenarios, attributable to platooning-induced aerodynamic drag reduction. CO$_2$ proxy emissions are higher under Cognitive as a direct consequence of its greater completed-trip volume, though per-kilometer emissions remain comparable across all scenarios.

\begin{figure}[htbp]
    \centering
    \includegraphics[width=0.82\columnwidth]{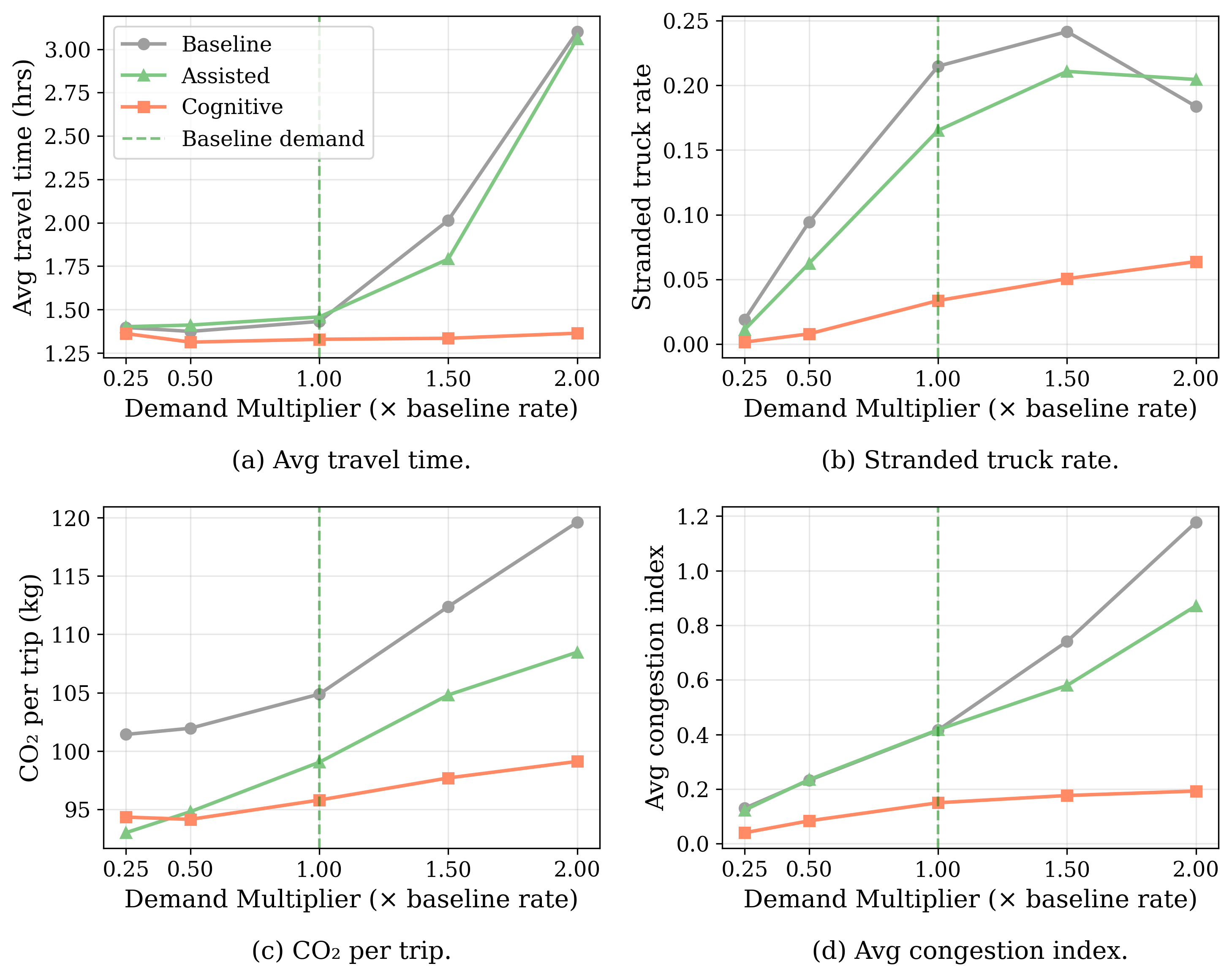}
    \caption{Performance across the three scenarios under varying origin demand levels.}
    \label{fig:fleet_sensitivity}
\end{figure}

\begin{figure}[htbp]
    \centering
    \includegraphics[width=0.82\columnwidth]{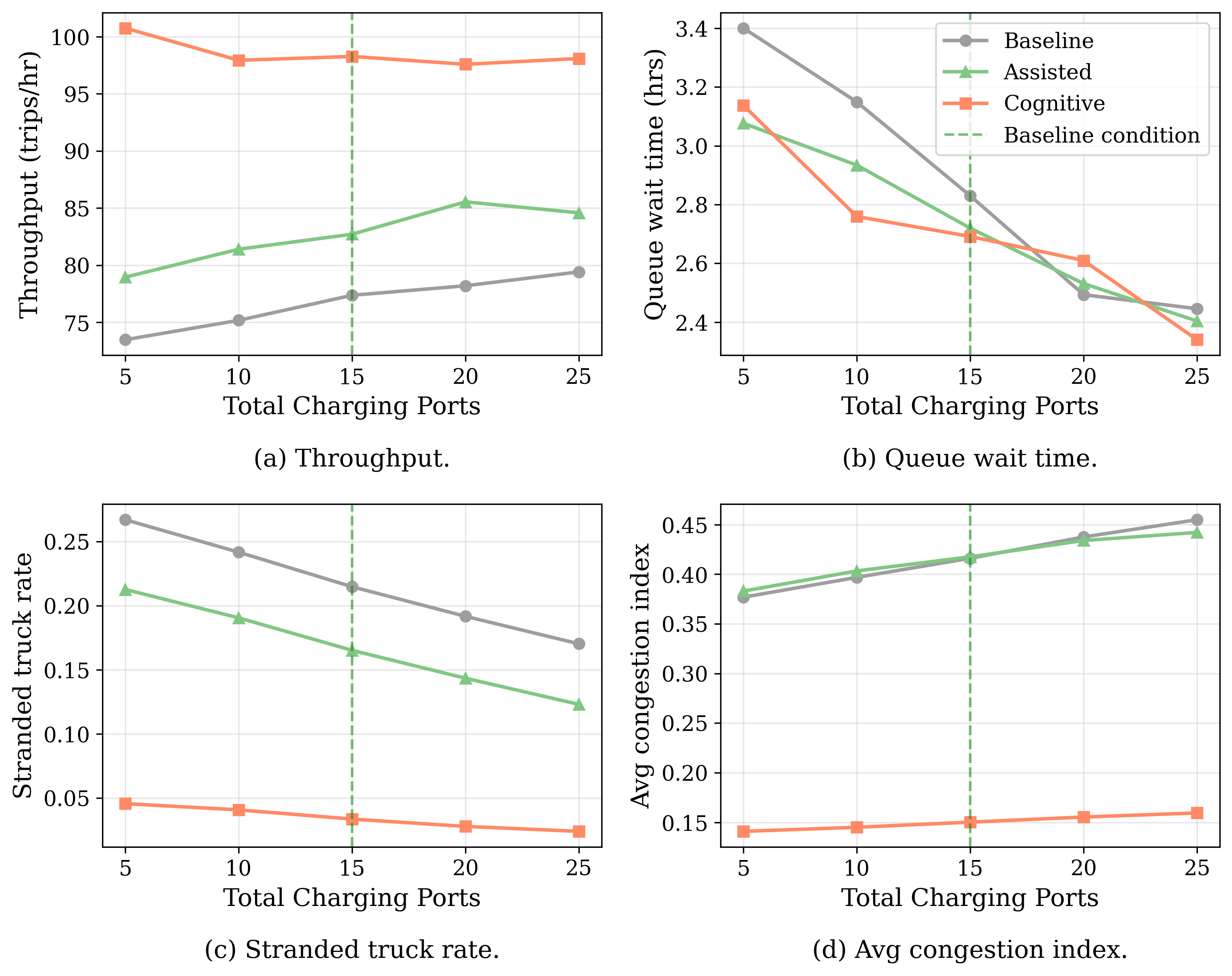}
    \caption{Performance across the three scenarios under varying numbers of corridor charging ports.}
    \label{fig:charging_sensitivity}
\end{figure}

\rev{To examine performance under different operating conditions, we conduct two sensitivity analyses. The demand sensitivity analysis (Figure~\ref{fig:fleet_sensitivity}) varies the origin terminal demand multiplier at five levels ranging from 0.25 to 2.0; the charging infrastructure analysis (Figure~\ref{fig:charging_sensitivity}) varies total corridor ports ranging from 5 to 25 in increments of 5 while holding demand fixed. Each condition is replicated five times.} 
\rev{Figure~\ref{fig:fleet_sensitivity} shows demand sensitivity across performance metrics. Panel~(a) shows that Baseline travel time rises steeply as demand increases, while Cognitive travel time remains stable at $\sim$1.34~hr (range 1.31--1.36~hr across all demand levels), reflecting sustained managed-lane and platoon effectiveness. Panel~(b) shows the stranded truck rate is substantially lower for Cognitive (0.001--0.064) than Baseline (0.019--0.242) across all demand levels, confirming MARL charging coordination reduces en-route energy depletion. Panel~(c) shows energy per trip is lowest under Cognitive at all demand levels. Panel~(d) confirms that the congestion advantage of Cognitive widens under high demand, where rule-based policies saturate.}


\rev{Figure~\ref{fig:charging_sensitivity} shows that queue wait time declines continuously from 5 to 25 ports across all scenarios, with the steepest improvement between 5 and 15 ports. Cognitive throughput remains stable across all port counts (98--101~trips/hr), whereas Baseline and Assisted show lower throughput at 5 ports (73.5 and 79.0~trips/hr, respectively, versus 77.4 and 82.7 at 15 ports). Stranded truck rates decline monotonically with additional ports for all scenarios, with Cognitive maintaining the lowest rates throughout. These results suggest diminishing returns to port additions beyond 15 ports, and that MARL coordination sustains performance across a wider range of charging infrastructure configurations than rule-based assignment.}

\begin{table}[htbp]
\small
\centering
\caption{Performance metrics across corridor scenarios.}
\label{tab:scenario_metrics}
\begin{tabular}{lccc}
\toprule
Metric & Baseline & Assisted & Cognitive \\
\midrule
Throughput (trips/hr) & 77.38 $\pm$ 37.38 & 82.72 $\pm$ 39.21 & 98.27 $\pm$ 47.09 \\
Avg Travel Time (hr) & 1.43 $\pm$ 0.03 & 1.46 $\pm$ 0.06 & 1.33 $\pm$ 0.07 \\
Energy (kWh/km) & 1.18 $\pm$ 0.00 & 1.10 $\pm$ 0.01 & 1.09 $\pm$ 0.01 \\
CO$_2$ Emissions (kg) & 101436.47 $\pm$ 48310.52 & 102404.27 $\pm$ 48048.79 & 117673.60 $\pm$ 54707.17 \\
Congestion Index & 0.42 $\pm$ 0.16 & 0.42 $\pm$ 0.17 & 0.15 $\pm$ 0.06 \\
\bottomrule
\end{tabular}
\end{table}

%
%

\subsection{Disruption Robustness} \label{subsec:disruption}

To evaluate robustness, we inject disruptions at 30\% of episode duration across four categories: accidents (single segment incident reducing capacity by 60\%), weather events (multiple segments affected with 30\% probability each and 30\% capacity reduction), station failures (one charging station becomes non-operational), and demand spikes (terminal demand doubles for remainder of episode).


\rev{Table~\ref{tab:disruption_outcomes} presents the robustness analysis across all three scenarios. All scenarios use the same controller trained under clean (no-disruption) conditions; disruptions are injected at evaluation time without retraining, constituting a zero-shot robustness assessment.}

\begin{table}[htbp]
\small
\centering
\caption{Robustness metrics under disruption scenarios.}
\label{tab:disruption_outcomes}
\begin{tabular}{lccccccccc}
\toprule
& \multicolumn{3}{c}{Baseline} & \multicolumn{3}{c}{Assisted} & \multicolumn{3}{c}{Cognitive} \\
\cmidrule(lr){2-4} \cmidrule(lr){5-7} \cmidrule(lr){8-10}
Disruption & Rob. & Thr. & TT$\uparrow$ & Rob. & Thr. & TT$\uparrow$ & Rob. & Thr. & TT$\uparrow$ \\
\midrule
Accident & 0.69 & 0.79 & 0.13 & 0.00 & 0.84 & 1.55 & 1.00 & 0.99 & -0.00 \\
Weather & 0.99 & 1.00 & 0.01 & 0.74 & 0.92 & 0.19 & 1.00 & 1.00 & 0.00 \\
Station Failure & 0.99 & 0.99 & -0.00 & 1.02 & 1.01 & -0.01 & 0.99 & 0.99 & 0.00 \\
Demand Spike & 0.52 & 0.92 & 0.43 & 0.70 & 1.14 & 0.39 & 1.31 & 1.32 & 0.00 \\
\bottomrule
\multicolumn{10}{l}{\footnotesize Rob. = Robustness score, Thr. = Throughput robustness, TT$\uparrow$ = Travel time increase ratio}
\end{tabular}
\end{table}



\subsection{Discussion} \label{subsec:discussion}
The results reveal scenario-differentiated performance rather than uniform monotonic improvement. Both Assisted and Cognitive deliver energy savings per kilometer ($-$7.5\% and $-$7.9\% vs.\ Baseline, respectively) through platooning, while primary advantages of the Cognitive scenario lie in throughput (\rev{+27.0\%}), travel time (\rev{$-$7.2\%}), and congestion reduction (\rev{$-$63.9\%}). These stem from qualitatively different mechanisms. V2X state sharing in Assisted enables better timing for charging stops and opportunistic platoon formation, whereas the RL controller in Cognitive actively manages policies to sustain flow at higher network utilization. Sensitivity analysis reinforces this interpretation, as throughput and congestion advantages in the Cognitive scenario widen most at high demand ($\geq 1.5\times$), where rule-based policies fail to adapt.



\rev{Under disruptions, Cognitive achieves the strongest overall resilience, maintaining near-full performance under three of four disruption types in zero-shot evaluation: accidents (1.00), weather (1.00), and demand spikes (1.31, where the policy exploits increased load to exceed undisrupted throughput). The sole exception is station failure, where Assisted (1.02) marginally outperforms both Cognitive (0.99) and Baseline (0.99) by redistributing charging demand across available ports without centralized coordination. Assisted collapses under accident conditions (0.00) due to a 155\% travel time spike from platoon congestion at the bottleneck, highlighting a vulnerability of rule-based platoon formation to capacity shocks. These zero-shot results suggest that RL-based corridor control generalizes robustly to most disruption types.}

For corridor planners, two implications follow from these preliminary results. First, V2X connectivity and assisted automation deliver measurable efficiency gains, particularly in energy consumption per kilometer ($-$7.5\% for Assisted), and represent a viable near-term deployment step. Second, RL/MARL control offers the greatest throughput scaling and congestion relief under certain conditions, motivating its deployment where demand regularly approaches or exceeds corridor capacity.

\section{CONCLUSIONS} \label{sec:conclusions}
This paper presented a three-layer, ABM of cognitive smart freight corridors in which physical infrastructure, V2X connectivity, and learning-based adaptive control are treated as an integrated corridor intelligent system. Experiments across scenarios confirm that RL/MARL control improves throughput and energy efficiency, and the Cognitive scenario demonstrates superior resilience under disruptions, supporting the argument that future corridor management should evolve from rule-based vehicle coordination toward adaptive, intelligent infrastructure capable of informing investment, operational policy, and governance decisions.

Several limitations motivate future extensions. First, the control framework could be broadened beyond platoon formation, managed lane control, and fixed station price adjustments to include dynamic rerouting, variable speed limits, and competitive pricing. Second, the current unidirectional network limits transferability to more realistic settings. Extending the model to bidirectional, multipath topologies and evaluating policy generalizability across different segment counts, demand patterns, and charging densities are therefore important next steps, including transfer across network configurations.

Further work is also needed to strengthen training and empirical validation. Reward ablations and extended training would clarify the contribution of each agent layer and address the convergence observed. Comparisons with established optimization benchmarks, additional replications, and paired statistical tests would provide stronger evidence beyond the directional findings reported here. Finally, calibration using real freight demand data and LLM-assisted policy interpretation could improve both fidelity and explainability.

\section*{ACKNOWLEDGMENTS}
This work was supported in part by the National Secretariat of Science, Technology, and Research (SENACYT) of Panama through its IFARHU-SENACYT Scholarship Program.


The views and opinions expressed in this article are those of the authors and do not necessarily reflect the position of Amazon.com, Inc. or its affiliates. This work was conducted independently of the authors' roles at their respective institutions.

\bibliographystyle{plainnat}
\bibliography{references}  






\end{document}